\documentclass[11pt]{article}
\usepackage[T1]{fontenc}
\usepackage[utf8]{inputenc}
\usepackage{amsmath,amssymb,amsthm}
\usepackage{graphicx}
\usepackage[margin=1in]{geometry}
\usepackage{booktabs}
\usepackage{multirow}
\usepackage[numbers,sort&compress]{natbib}
\usepackage[colorlinks=true,linkcolor=blue,citecolor=blue,urlcolor=blue]{hyperref}
\usepackage{authblk}
\usepackage{xspace}

\newcommand{\A}{\ensuremath{A}\xspace}
\newcommand{\M}{\ensuremath{M}\xspace}
\newcommand{\Adev}{\ensuremath{A_{\mathrm{dev}}}\xspace}
\newcommand{\Kdev}{\ensuremath{K_{\mathrm{dev}}}\xspace}
\newcommand{\ksup}{\ensuremath{k_{\sup}}\xspace}
\newcommand{\supdev}{\ensuremath{\sup_k\|A_{\mathrm{dev}}^k\|_2}\xspace}
\newcommand{\eps}{\ensuremath{\varepsilon}\xspace}
\newcommand{\rhoe}{\ensuremath{\rho_\eps}\xspace}
\newcommand{\Rope}{RoPE\xspace}
\newtheorem{prop}{Proposition}

\title{\textbf{Transient Reserves, Sink Dampers, and the Failure of\\
Eigenvalue Reasoning in the Attention Propagator}}
\author[1]{Li Hengyu}
\affil[1]{Institute for Solid State Physics, The University of Tokyo\authorcr
\texttt{lihengyu@issp.u-tokyo.ac.jp}}
\date{}

\begin{document}
\maketitle

\begin{abstract}
The attention matrix of a causal transformer is row-stochastic, iterated over depth, and non-normal by
construction. For non-normal operators, eigenvalues control only asymptotic behavior; finite-depth behavior
is controlled by resolvent quantities such as pseudospectra and Kreiss constants. We test, under
pre-registered criteria, whether this resolvent view predicts anything about trained transformers that
eigenvalues miss. Two structural facts organize the analysis: the mask pins the Kreiss constant of every
causal stochastic matrix at $\sqrt{n}$, and deflating the mask-forced Perron projector factorizes the depth
deviation dynamics exactly into a product of deflated operators. Across GPT-2, Pythia-410m, and Llama-3-8B,
learned non-normality proves to be signed. A routing minority carries excess transient reserve that tracks
previous-token function and doubles when induction heads engage, while the sink majority is suppressed below
matched shuffle nulls, so that attention sinks act as transient dampers. On depth products, eigenvalue
predictions of surviving deviations err by seven to eleven orders of magnitude, an error absent in matched
nulls. Checkpoint censuses date this organization to a consolidation phase after circuit formation, and a
clamping intervention on Llama-3-8B establishes a causal chain from three massive-activation dimensions
through sink attention to transient damping; LayerNorm models implement the same functions elsewhere. A
cross-validated contest concludes that resolvent features are required for depth-transient persistence and
routing-head identity, and that no single-operator summary of any kind predicts per-head causal criticality.
\end{abstract}

% ================================================================ 1
\section{Introduction}
\label{sec:intro}

A causal transformer applies, at every layer and head, a row-stochastic attention matrix
$\A\in\mathbb{R}^{n\times n}$ to the token stream. Whatever else the architecture does, the composition of
these matrices over depth is its token-mixing skeleton: information moves through the products
$\A^{(L)}\cdots\A^{(1)}$. It is natural, and common, to reason about such objects through eigenvalues, using
spectral gaps for mixing rates, dominant eigenvectors for stationary structure, and spectral radii for
growth \citep{dong2021attention,naderi2024gap}.

For normal operators this reasoning is sound. The attention matrix is not normal; the causal mask makes it
lower-triangular, which is a maximally non-normal structure. For non-normal operators the eigenvalues govern
only the limit $k\to\infty$ of $\|\A^k\|$, whereas finite-horizon behavior, the only behavior a finite-depth
network exhibits, is governed by resolvent quantities: the \eps-pseudospectrum
$\Lambda_\eps(\A)=\{z:\ \sigma_{\min}(zI-\A)<\eps\}$ and the Kreiss constant
$K(\A)=\sup_{|z|>1}(|z|-1)\,\|(zI-\A)^{-1}\|_2$, which brackets the transient through
$K(\A)\le\sup_k\|\A^k\|\le enK(\A)$ \citep{trefethen2005spectra}. In fluid mechanics, this gap between
eigenvalue stability and transient amplification resolved a long-standing mismatch between linear theory and
observed transition to turbulence \citep{trefethen2005spectra}. In recurrent networks and neural circuits,
non-normal transients are a recognized computational resource \citep{kerg2019nnrnn,murphy2009balanced,
goldman2009memory}. Recent deep-learning applications of pseudospectral tools target optimizer dynamics
\citep{ghosh2026nonnormal,ariq2026pseudospectral}. The forward-pass attention operator itself, although it is
iterated over depth, non-normal by construction, and central to interpretability, has not been examined
through this lens.

This paper does so under a discipline fixed in advance. The claim under test is that for the attention
propagator, eigenvalue analysis is insufficient or misleading, and the pseudospectrum predicts behavioral
structure that eigenvalues miss. We pre-registered the three admissible verdicts (yes, no, or only for a
characterized subdomain), the decisive experiment (a cross-validated contest between eigenvalue-side and
resolvent-side features, Section~\ref{sec:decisive}) together with its verdict rule, and a commitment to
reporting negative results at full prominence. Two methodological hazards make this discipline necessary.
First, the mask supplies dramatic non-normality for free, so any claim about learned structure must be
measured as excess over mask-matched nulls. Second, some resolvent quantities are related to candidate
observables by theorems rather than by empirical facts about transformers; the one place where this
tautology arises is flagged explicitly in Section~\ref{sec:decisive}.

\paragraph{Contributions.}
\begin{enumerate}\itemsep2pt
\item \textbf{Mask-trivia results and a deflation identity} (\S\ref{sec:setup}). The Kreiss constant of every
causal stochastic matrix is pinned at $\sqrt{n}$ by the mask, so the raw constant carries no information
about learning. The mask-forced Perron projector $P=\mathbf{1}e_0^\top$ yields the exact splitting
$\A^k=P+\Adev^k$ and, for layer-mean stacks, $\Pi_L-P=\prod_l\bar A^{(l)}_{\mathrm{dev}}$. All subsequent
analysis is carried out on the deflated operator against two mask-structure nulls.
\item \textbf{A signed census of learned non-normality} (\S\ref{sec:census}). Across three architectures, a
routing minority carries excess transient reserve (the correlation between Kreiss excess and previous-token
score is $+0.54$ to $+0.60$ in all models), including heads whose deviation dynamics grows to an interior
peak with the Kreiss bound nearly attained. The sink majority is suppressed below the mass-matched null,
increasingly so with scale (86\% of GPT-2 heads; 99.7\% of Llama heads). Sinks are transient dampers.
\item \textbf{A learned, depth-growing failure of eigenvalue reasoning} (\S\ref{sec:depth}). On real depth
products, the eigenvalue heuristic underpredicts surviving deviations by $10^{7}$ to $10^{11}$, while
matched nulls are off by a small factor. Perron absorption is accelerated by training.
\item \textbf{State-conditionality} (\S\ref{sec:awake}). Induction heads behave as dormant sinks whose
Kreiss constant doubles when the input engages them, an effect controlled for mass profile and replicated in
all three models.
\item \textbf{Training-time consolidation} (\S\ref{sec:training}). Thirteen-checkpoint censuses on
Pythia-410m and Pythia-160m show the propagator-side suppression locking in after circuit formation and
crossing half-depth at the same checkpoint as the weight-space spectral suppression reported by the
companion paper, with the sink pattern never lagging its damping function.
\item \textbf{A causal, architecture-conditional mechanism} (\S\ref{sec:massive}). Clamping three
massive-activation dimensions on Llama-3-8B deletes the sink and releases the damper ($\Kdev$ rises from
$1.4$ to $7.4$), whereas LayerNorm models implement both functions elsewhere. The hypothesis that massive
activations are themselves a non-normal transient overshoot is rejected on timing in all three models.
\item \textbf{The decisive contest} (\S\ref{sec:decisive}). Resolvent features win robustly for
depth-transient persistence (three of three models) and routing identity (two of three); every feature
family fails for induction identity and for per-head causal criticality, which marks the boundary of
single-operator summaries.
\end{enumerate}

A companion paper \citep{li2026qk} analyzes the static scoring operator $\M=W_Q^{\top}W_K$ of the same models
with eigenvalue-level tools; Section~\ref{sec:discussion} states the division of labor that the two studies
jointly establish.

% ================================================================ 2
\section{Setup: the propagator, the mask, and what the mask gives for free}
\label{sec:setup}

\paragraph{Objects.} For a model with $L$ layers and $H$ heads we capture post-softmax, post-mask attention
matrices $\A^{(l,h)}(x)\in\mathbb{R}^{n\times n}$ on a fixed corpus (pile-10k \citep{gao2020pile} rows of
$n{=}128$ tokens, BOS-prefixed). These are per-input objects and are analyzed distributionally. The models
are GPT-2 small (12 layers $\times$ 12 heads, LayerNorm, learned absolute positions), Pythia-410m
(24 $\times$ 16, LayerNorm, partial \Rope), and Llama-3-8B (32 $\times$ 32, RMSNorm, full \Rope, grouped-query
attention). All spectral computation is in float64, and rows are renormalized after float32 or bfloat16
capture.

\paragraph{What the mask forces.} Causality makes \A lower-triangular, so its eigenvalues are its diagonal
and row $0$ is forced to $e_0^\top$. The Perron pair is therefore fixed by the mask alone, with right vector
$\mathbf{1}$ and left vector $e_0$, for every causal stochastic matrix, learned or random. Two consequences
organize what follows.

\begin{prop}[Mask-trivial Kreiss constant]\label{prop:sqrtn}
Let $\A\in\mathbb{R}^{n\times n}$ be causal (lower-triangular) and row-stochastic with $A_{ii}<1$ for $i>0$.
Then $P=\mathbf{1}e_0^\top$ satisfies $AP=PA=P^2=P$ and $\|P\|_2=\sqrt{n}$, and since
$(zI-\A)^{-1}=P/(z-1)+O(1)$ as $z\to1^{+}$,
\[
K(\A)\;=\;\sup_{|z|>1}(|z|-1)\,\|(zI-\A)^{-1}\|_2\;\ge\;\|P\|_2\;=\;\sqrt{n}.
\]
\end{prop}

Empirically the bound saturates universally: $K(\A)=11.314=\sqrt{128}$ to three decimals for every real and
null matrix in this study, including heads whose deflated constant exceeds $\sqrt n$ (for example
$\Kdev=15.7$), where the Perron pole and the deviation resolvent interfere rather than add. In either case
the raw Kreiss constant of causal attention measures the mask, not the learned dynamics, and the informative
object is the deflated operator.

\begin{prop}[Exact Perron deflation]\label{prop:deflate}
With $\Adev := \A - P$, one has $\A^k = P + \Adev^k$ for all $k\ge1$. For layer stacks of head-mean matrices
$\bar A^{(l)}$, each of which shares the mask-forced $P$, the depth products satisfy
\[
\Pi_L - P \;=\; \bigl(\bar A^{(L)}-P\bigr)\cdots\bigl(\bar A^{(1)}-P\bigr)
\;=\; \prod\nolimits_l \bar A^{(l)}_{\mathrm{dev}} .
\]
\end{prop}

The deviation-from-stationarity dynamics, which is the object relevant to over-smoothing, rank collapse, and
information survival over depth, is therefore exactly the (product of) deflated operator(s). All transient
metrics below are computed on \Adev: the Kreiss constant $\Kdev=K(\Adev)$, the pseudospectral radius
$\rhoe(\Adev)$ at $\eps\in\{10^{-1},10^{-2},10^{-3}\}$, the measured transient \supdev with its argmax
\ksup, and structural covariates (BOS-column mass, diagonal mass, row entropy, and the Dobrushin coefficient
$\tau$).

\paragraph{Computation.} We evaluate $\sigma_{\min}(zI-\A)$ by one complex Schur factorization per matrix
followed by inverse-power iteration with batched triangular solves, at cost $O(n^2)$ per grid point, with
residual-based convergence and analytic floors $\rhoe\ge\rho+\eps$ that guard isolated-eigenvalue balls
against under-resolution by finite grids \citep{trefethen1999computation}. The implementation reproduces
textbook pseudospectra (for the Grcar matrix, and for Jordan blocks, where \rhoe follows the $\eps^{1/n}$
law to 1\%), matches dense SVD to $10^{-6}$ relative error, and passes a regression suite that includes
numerical verification of both propositions. Census-scale settings were validated against full settings
(Kreiss constants agree to 0.0\%, \rhoe to 0.01\%). The study comprises roughly $3.5\times10^{5}$ matrix
analyses; the larger censuses ran on a 128-core cluster node in batches of at most 28 minutes.

\paragraph{Nulls.} Because the mask supplies non-normality for free, every quantity is reported as excess
over per-matrix nulls. The primary null is a row permutation: each row's attention values are permuted
within its causal support, which preserves every row's mass profile exactly (peakedness and entropy) while
destroying learned column and diagonal structure. A pool of Dirichlet lower-triangular stochastic matrices
serves as a generic reference. On the eigenvalue side, the matched-normal control is analytic: a normal
matrix with the same spectrum has $K=1$ and $\rhoe=\rho+\eps$ exactly, so $\Kdev>1$ already constitutes the
comparison against an eigenvalue-matched normal operator.

% ================================================================ 3
\section{A signed census: routing reserves and sink dampers}
\label{sec:census}

For every head we compute the deflated battery on 8 to 32 corpus matrices, each paired with its
row-permutation null, using per-head Wilcoxon tests over sequences and Benjamini--Hochberg correction over
heads.

\begin{table}[t]
\centering\small
\begin{tabular}{lccc}
\toprule
 & GPT-2 (144 heads) & Pythia-410m (384) & Llama-3-8B (1024)\\
\midrule
FDR-significant vs.\ row-perm null & 137/144 & 368/384 & 1023/1024\\
\quad of which above (excess) & 12 & 38 & 2\\
\quad of which below (suppressed) & 125 & 330 & 1021\\
median $\log_2$ \Kdev excess & $-1.55$ & $-1.68$ & $-2.20$\\
\midrule
$\rho$(\Kdev excess, prev-token score) & $+0.55$ & $+0.60$ & $+0.54$\\
$\rho$(\rhoe excess, prev) & $+0.62$ & $+0.57$ & $+0.58$\\
$\rho$(BOS mass, prev) & $-0.52$ & $-0.63$ & $-0.50$\\
$\rho$(\Kdev excess, induction score) & $-0.52$ & $-0.33$ & $+0.02$\\
\bottomrule
\end{tabular}
\caption{The signed census. Learned attention is predominantly less non-normal than its mass-matched null
(the sink majority, whose suppression deepens with scale), while a routing minority carries genuine excess.
Head-class scores (previous-token, induction) are behavioral and come from the companion paper's probe
taxonomy. All correlations shown survive BH-FDR at $q{=}0.001$ except the Llama induction entry, which is
not significant.}
\label{tab:census}
\end{table}

\begin{figure}[t]
\centering
\includegraphics[width=.49\linewidth]{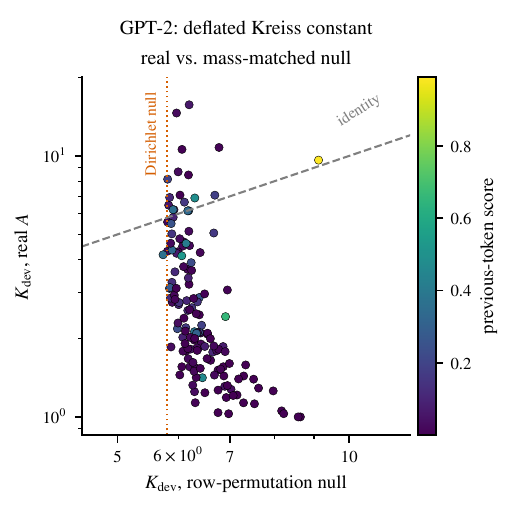}\hfill
\includegraphics[width=.49\linewidth]{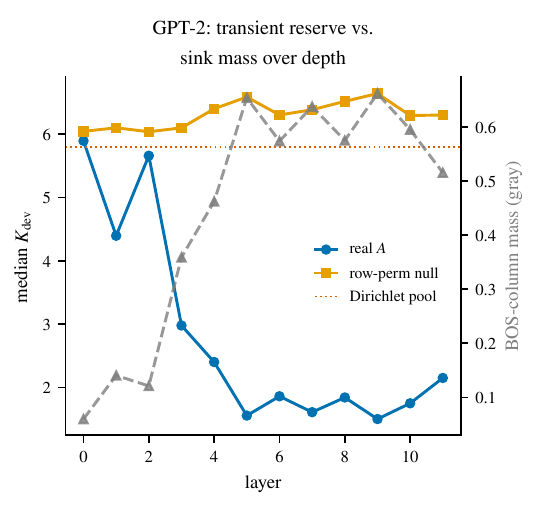}
\caption{Left: per-head median \Kdev, real versus row-permutation null (GPT-2, log--log axes). The nulls
concentrate in a narrow band near 6 while real heads spread across $[1,15.7]$; previous-token-like heads
(bright) lie on or above the diagonal, and sink-dominated heads collapse to the analytic normal floor
$\Kdev=1$. Right: the damper over depth. The median real \Kdev tracks the null in layers 0--2 and then
collapses to between 1.5 and 2 exactly as the BOS sink mass (gray) rises; sinks purchase depth stability by
suppressing transient reserve.}
\label{fig:census}
\end{figure}

Three structures stand out (Table~\ref{tab:census}, Fig.~\ref{fig:census}).

First, the sink majority is suppressed below chance. In GPT-2, Pythia, and Llama respectively, 86\%, 86\%,
and 99.7\% of heads sit significantly below their mass-matched null. The strongest sinks, such as GPT-2's
dormant induction heads 5.1, 7.2, and 8.1 with BOS mass between $0.79$ and $0.97$, reach the analytic normal
floor $\Kdev=1.00$ with measured transients $\supdev<3$. Attention sinks are thus not merely attractors of
attention mass; at the operator level they are transient dampers, and the damping deepens with model scale
(median excess $-1.55$ in GPT-2 versus $-2.20$ in Llama).

Second, a routing minority carries learned excess, including genuine interior amplification. The excess
heads are early-layer positional heads together with two pronounced outliers, GPT-2 heads L1H10 and L11H8
($\Kdev=14.6$ and $15.7$). For these two, the deviation dynamics $\|\Adev^k\|$ grows with $k$ to an interior
peak at $\ksup=19$ and $64$, with $\supdev=15.6$ and $15.9$, so the Kreiss lower bound is nearly attained;
every null has $\ksup\equiv1$. The dominant transient mode is interpretable: the top singular pair of
$\Adev^{\ksup}$ reads the contrast between BOS and the first content token ($v\approx(e_0\pm e_1)/\sqrt2$)
and writes it nearly uniformly across positions ($|u|\approx n^{-1/2}$), which amounts to a broadcast
amplifier of early-context contrast (Appendix~\ref{app:amplifier}).

Third, an honest null result. The canonical previous-token head 4.11 (previous-token score $0.99$) has a
high $\Kdev$ of $9.6$, and its pseudospectrum forms a ring near $|z|=1$, the signature of a truncated
Toeplitz shift whose eigenvalues, the mask-forced Perron value at $1$ aside, cluster near zero and are
uninformative \citep{trefethen2005spectra}. Its
row-permutation null, however, matches both statistics ($K=9.1$; $\rho_{10^{-2}}$ of 1.082 versus 1.096).
Scalar resolvent levels of near-deterministic routing are therefore explained by row peakedness together
with causality; what no null reproduces is the interior growth described above. The classical mixing bound
also fails at exactly this head: its Dobrushin coefficient is $\tau=1.000$, which is vacuous, while its
measured deviation residual at depth $n$ is $0.34$. Norm- and eigenvalue-side rate predictors are
uninformative precisely where the resolvent view is needed.

% ================================================================ 4
\section{Depth products: a learned failure of the eigenvalue heuristic}
\label{sec:depth}

By Proposition~\ref{prop:deflate}, the depth deviation-transient of the head-mean stack is
$D_L=\|\prod_{l\le L}\bar A^{(l)}_{\mathrm{dev}}\|_2$. We compare it with the eigenvalue heuristic
$E_L=\prod_l\rho(\bar A^{(l)}_{\mathrm{dev}})$, computed from spectral radii (exact diagonals for triangular
matrices), and with the submultiplicative bound $N_L=\prod_l\|\bar A^{(l)}_{\mathrm{dev}}\|_2$, per input
sequence, for the real stack, for a per-layer row-permutation null, and for the rollout variant
$\tfrac12(I+\bar A)$ of \citet{abnar2020rollout}.

\begin{table}[t]
\centering\small
\begin{tabular}{lcccc}
\toprule
model (depth) & $D_{L_{\max}}$ & $E_{L_{\max}}$ (eigen) & $D/E$ real & $D/E$ null\\
\midrule
GPT-2 (12) & $5.1\times10^{-4}$ & $1.8\times10^{-11}$ & $\mathbf{2.9\times10^{7}}$ & $4.8$\\
Pythia-410m (24) & $2.6\times10^{-11}$ & $3.5\times10^{-18}$ & $\mathbf{7.4\times10^{6}}$ & $7.5\times10^{-4}$\\
Llama-3-8B (32) & $9.7\times10^{-22}$ & $7.5\times10^{-33}$ & $\mathbf{1.3\times10^{11}}$ & $5.9\times10^{-8}$\\
\bottomrule
\end{tabular}
\caption{Eigenvalue reasoning about depth fails on trained stacks, and the failure is learned. Entries are
medians over sequences at final depth. Real stacks keep deviations alive $10^{7}$ to $10^{11}$ times longer
than the eigenvalue prediction, whereas the mass-matched null is off by a small factor in GPT-2 and its error
even changes sign with depth. The norm bound overshoots by 9 to 33 orders of magnitude (not shown). The same
orders of magnitude hold on an induction-probe corpus.}
\label{tab:depth}
\end{table}

\begin{figure}[t]
\centering
\includegraphics[width=.9\linewidth]{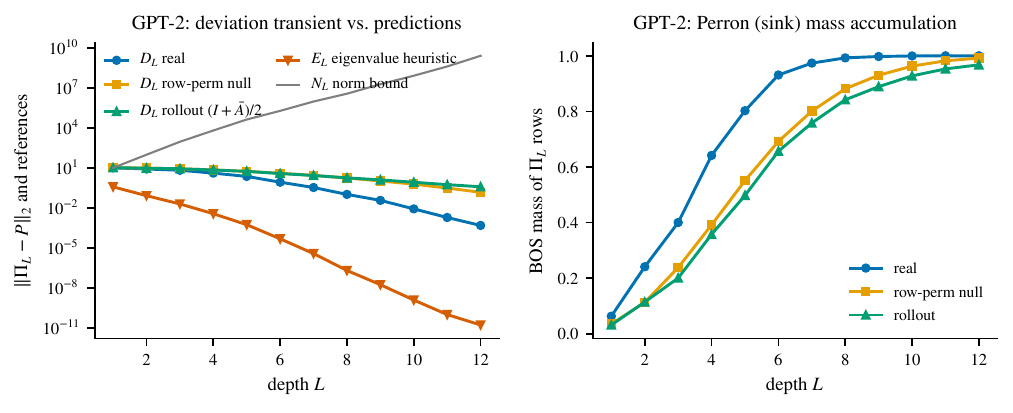}
\caption{Depth deviation-transient for GPT-2. Left: the measured $D_L$ (blue) against the eigenvalue
heuristic $E_L$ (red, which collapses to $10^{-11}$) and the norm bound (gray, which grows to $10^{9}$); the
measured curve lies many orders away from both. The row-permutation null (orange) decays more slowly in
absolute terms, yet its eigenvalue heuristic is nearly calibrated. Right: accumulation of Perron (sink) mass.
The real stack absorbs to BOS by roughly layer 8, far earlier than the null, so Perron absorption is
accelerated by training.}
\label{fig:depth}
\end{figure}

Table~\ref{tab:depth} and Fig.~\ref{fig:depth} support three readings. The eigenvalue heuristic is not
merely loose but qualitatively wrong: trained stacks concentrate diagonal mass in few heads, the head-mean
diagonals are therefore small, and $E_L$ collapses, while the actual mixing is slower by many orders of
magnitude. The failure is learned: for the mask-null stack the same heuristic is nearly calibrated at GPT-2
depth ($D/E=4.8$), and the gap between real and null grows from about $10^{7}$ to about $10^{18}$ across the
depth range studied. Finally, the residual-path rollout keeps $D_{12}=0.41$, so no over-smoothing
catastrophe occurs once skip connections are included, in line with the counteraction result of
\citet{dong2021attention}, while still exceeding its eigenvalue heuristic by a factor of about 300. We state
the scope plainly: $E_L$ is a heuristic rather than a bound for products of distinct operators. The point is
that the default mental model of what the spectrum says about depth errs by many orders of magnitude on
trained networks, and errs in the direction of longer survival.

% ================================================================ 5
\section{The reserve is state-conditional: dormant sinks wake into routers}
\label{sec:awake}

On generic text, induction heads read as pure sinks, parked on BOS with $\Kdev\approx1$
\citep{guo2024activedormant}. We repeated the full census on induction-triggering probes of the form BOS
$+\,r+r$ with random $r$. The probes behave as intended: induction-stripe mass reaches $0.84$ to $0.93$ on
the second copy while previous-token stripes remain at $0.99$. Each state is paired with its own
row-permutation null, so changes in mass profile are controlled.

\begin{figure}[t]
\centering
\includegraphics[width=.62\linewidth]{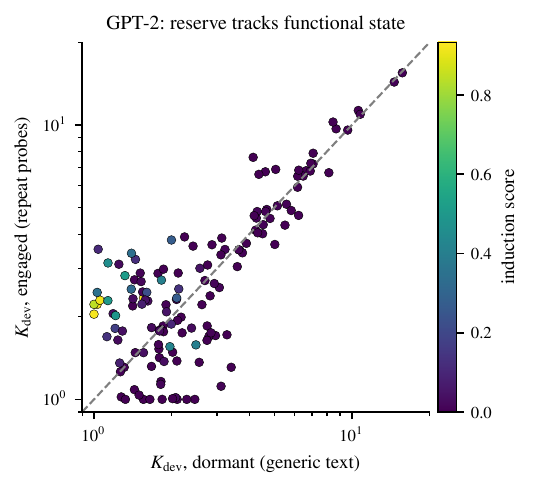}
\caption{Transient reserve tracks functional state (GPT-2). Induction heads (bright) leave the dormant sink
corner near $\Kdev=1$ and land above the identity when engaged, at $\Kdev$ between 2.0 and 2.8. Persistent
routing heads sit on the diagonal. The shift is an excess gain over state-matched nulls and therefore
reflects added routing structure rather than a changed row profile.}
\label{fig:awake}
\end{figure}

Engagement doubles the Kreiss constant of induction heads (in GPT-2, \Kdev moves from the range $1.0$--$1.6$
to $2.0$--$2.8$, with excess gains of $+0.6$ to $+1.2$ bits over state-matched nulls). The effect is
class-specific: the correlation of the excess change with induction score is $+0.43$ ($p=10^{-7}$), $+0.44$
($p=10^{-19}$), and $+0.18$ ($p=8\times10^{-9}$) across GPT-2, Pythia, and Llama; duplicate-token heads also
engage; and previous-token heads, which are already active, do not move (correlations $-0.25$ to $-0.28$).
At the operator level, head engagement is a transition between damper and router: the same head that damps
transients on generic text becomes a shift-like transporter when its trigger appears, with the first copy
still parked on BOS and the second copy shifted back by the repeat period. Together with the companion
paper, this explains why static weight fingerprints cannot single out induction heads; their distinguishing
operator property is input-conditional by construction.

% ================================================================ 6
\section{Training time: the damper consolidates after formation}
\label{sec:training}

When is this organization learned? We ran the identical battery over 13 Pythia checkpoints (steps $0$ to
$143$k, corresponding to $0$ to $300$B tokens) for Pythia-410m and for a Pythia-160m replicate, adding about
$2.2\times10^{5}$ matrix analyses. The pipeline is bit-compatible with the main census: at the final
checkpoint the per-matrix Kreiss values reproduce the main-census values with Spearman correlation
$1.000000$.

\begin{figure}[t]
\centering
\includegraphics[width=.49\linewidth]{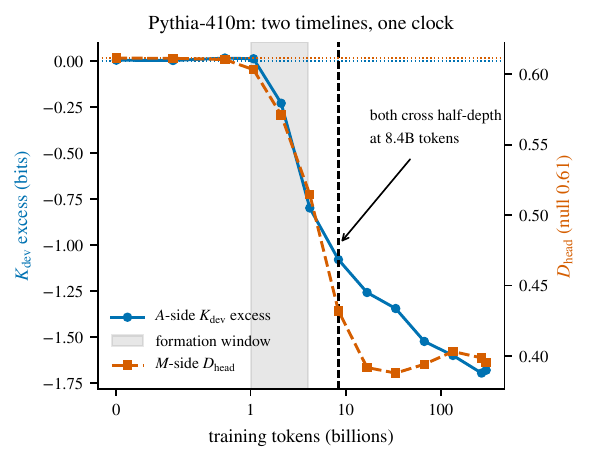}\hfill
\includegraphics[width=.49\linewidth]{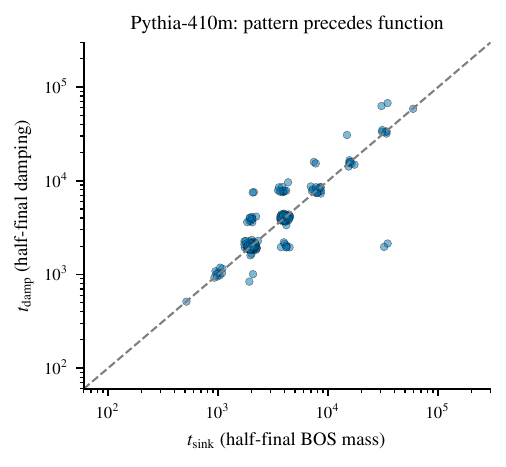}
\caption{Training-time consolidation on Pythia-410m. Left: the population median \Kdev excess (propagator
side, blue) and the companion paper's weight-space $D_{\mathrm{head}}$ suppression (red) on a shared token
axis. Both sit at their nulls through 1.1B tokens, begin to decline in the tail of the circuit-formation
window (shaded), and cross half of their final depth at the same checkpoint, 8.4B tokens. The propagator
side keeps deepening to 300B, while the weight-side all-head median saturates by roughly 34B; the latter is
a mixture statistic, and the companion reports its induction subgroup deepening well past that point.
Right: per-head timing. The sink attention pattern forms before or together with its damping in 94.8\% of
damped sink heads, and no head in either model shows damping at baseline sink mass.}
\label{fig:training}
\end{figure}

Figure~\ref{fig:training} supports three statements. First, suppression is a consolidation process that
follows circuit formation. The population median \Kdev excess sits at the null through step 512 (1.1B
tokens), moves during the second half of the formation window (14\% of final depth at 2.1B tokens), and
crosses half of its final depth at 8.4B tokens, well after formation. Second, it shares a clock with the
weight-space suppression. The companion paper's population $D_{\mathrm{head}}$ trajectory crosses its own
half-depth at the same checkpoint (trajectory Spearman $+0.79$, $p=1.5\times10^{-3}$), and in the 160m
replicate both cross at the same, one notch earlier, checkpoint of 4.2B tokens. The within-family trend also
reproduces the cross-model one, with final suppression depth $-0.62$ at 160m versus $-1.68$ at 410m. The
population medians conceal a common functional stratification: on the propagator side, with head classes as
in Section~\ref{sec:awake}, induction-class heads damp deepest (median excess $-2.81$ by 300B tokens),
previous-token heads retain positive reserve throughout training (about $+0.5$ from 2B tokens onward), and
the remainder settles in between ($-1.6$), mirroring the subgroup structure that the companion reports in
weight space. Third, the pattern precedes the function. With onsets defined by half-of-final crossings, the
sink attention pattern forms before or together with its \Kdev collapse in 202 of 213 damped sink heads
(94.8\%, above the pre-registered 90\% bar). At 160m the same readout gives 15 of 19 (79\%), below the bar,
and we report this as a formal miss, although all four exceptions are curvature races between smooth,
concurrently moving trajectories. A secondary readout with absolute onset thresholds, frozen before
re-analysis but post hoc with respect to the census data and reported as such, sharpens rather than rescues
the picture: 410m passes at 207 of 213 (97.2\%), 160m misses by one head (17 of 19, 89.5\%), and the three
heads that formally violate the no-damping-without-sink clause at 160m turn out to be final-layer
anchor-concentration heads, with near-one-hot rows (normalized entropy $0.01$ to $0.02$) concentrated on a
few recurring non-BOS columns, which a BOS-based sink definition cannot detect. Across both models, no head
at any checkpoint (0 of 522) is deeply damped without a concentration pattern of some kind. The general law
is concentration accompanied by damping, with BOS-sink damping as its large-model special case; what fails
at small scale is the BOS-based operationalization, not the ordering claim.

% ================================================================ 7
\section{Massive activations: timing, and a causal arrow that depends on architecture}
\label{sec:massive}

Massive activations, a handful of residual-stream coordinates with $10^{3}$ to $10^{4}$ times the typical
magnitude, are known to accompany attention sinks \citep{sun2024massive,xiao2024streamingllm,
gu2024sinkformation}. A natural conjecture, which we examined and now reject, is that the massive burst is
itself a non-normal transient overshoot of the attention dynamics.

Timing rejects the overshoot direction in all three models. Comparing like-for-like onsets (the largest jump
of the maximal residual magnitude at BOS versus the largest jump of per-layer BOS attention), massive
activations lead the sink by two layers in GPT-2 and one layer in Pythia, and are simultaneous with it in
Llama; they never lag. Magnitude and space agree with this verdict: measured attention-side transients are
bounded by $\Kdev\le16$ and act in position space with flat write profiles, whereas massive activations are
events of order $3000$ locked to specific residual dimensions (GPT-2: 447, 378, 138; Pythia: 130, 357, 966;
Llama: 788, 1384, 4062). Position-level co-location is architecture-dependent (the rank correlation between
stationary mass and per-position massive magnitude is $-0.11$ on GPT-2 but $+0.29$ and $+0.34$ on Pythia and
Llama), which is consistent with the bias-creates-sink direction of \citet{sun2024massive} and not with an
overshoot.

\begin{figure}[t]
\centering
\includegraphics[width=.95\linewidth]{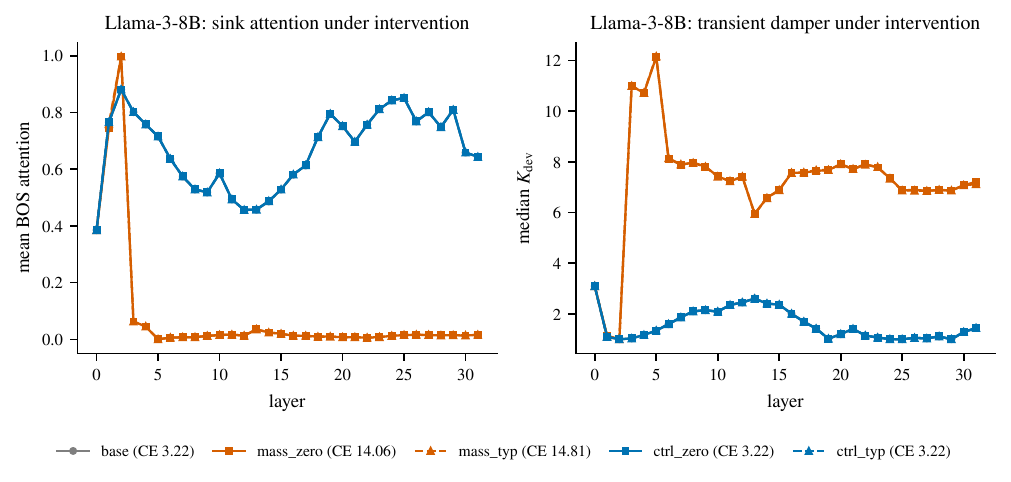}
\caption{The causal test on Llama-3-8B. The three massive dimensions are clamped at every residual boundary
(red) against three random dimensions (blue). BOS attention collapses from layer 3 onward ($0.675$ to
$0.049$, left) and the Kreiss damper releases, with \Kdev jumping to 11--12 at layers 3--5 and settling near
$7.4$, approximately the mask-null level (right). Controls are indistinguishable from baseline to three
decimals.}
\label{fig:intervention}
\end{figure}

An intervention establishes the causal arrow where the architecture permits. We clamped the three massive
dimensions, either to zero or to their per-layer median, at every residual boundary, and measured attention
structure, \Kdev, and loss against random-dimension controls (Fig.~\ref{fig:intervention}). On Llama-3-8B
the effect is surgical. BOS attention collapses ($0.675$ to $0.049$) while self- and previous-token stripes
are untouched ($0.047$ to $0.044$ and $0.046$ to $0.040$) and row entropy rises from $1.37$ to $3.84$ nats;
the channel carries specifically the sink component, and the freed mass diffuses rather than relocating. At
the same time the damper releases, with \Kdev rising from $1.41$ to $7.44$. On GPT-2 and Pythia the same
intervention leaves both the sink pattern and its damping intact ($0.547$ to $0.550$ with \Kdev $1.94$ to
$2.40$; $0.640$ to $0.713$ with \Kdev $1.63$ to $1.14$), although it is equally destructive functionally
(cross-entropy rises from $3.8$, $3.4$, and $3.2$ to $8.3$, $9.2$, and $14.1$ respectively, while
random-dimension controls cost at most $0.18$). The mechanism is therefore architecture-conditional. Models
with RMSNorm and full \Rope route the sink bias through a narrow and interceptable dimension channel, while
LayerNorm models implement sink and damping elsewhere, mirroring at the level of implementation the
companion paper's architecture-conditional verdict at the level of algebra. Caveats: whole-stream clamping
is stronger than the pointwise interventions of \citet{sun2024massive}, and loss increases in all models;
the sample is eight sequences; only the top three dimensions were clamped.

% ================================================================ 8
\section{The decisive contest}
\label{sec:decisive}

The pre-registered endpoint is a cross-validated prediction contest between three feature families.
Family E contains spectral and classical quantities: the deflated diagonal and spectral radius, the
Dobrushin coefficient, operator norms, BOS and diagonal mass, entropy, and the companion paper's weight-space
eigenvalue features. Family P contains resolvent quantities: \Kdev, \rhoe with its excesses, \ksup, the
Henrici departure, eigenvector conditioning, and the weight-space non-normality scalars. The third family is
their union. The model is ridge regression with inner-loop regularization, grouped cross-validation over
heads (so the dynamics contest generalizes to unseen heads), 10 to 40 repeats, and two feature
preprocessings (raw-standardized and rank-normal). Support is declared only when the pre-registered
criterion, namely $R^2(\mathrm{P})>R^2(\mathrm{E})$ with a confidence interval excluding zero or
$\Delta R^2(\mathrm{E}\to\mathrm{E\cup P})$ with a confidence interval excluding zero, holds under both
preprocessings.

\begin{table}[t]
\centering\small
\begin{tabular}{lccc}
\toprule
observable & GPT-2 & Pythia-410m & Llama-3-8B\\
\midrule
depth-transient persistence & P$\checkmark$ (incr.) & P$\checkmark$ (incr.) & \textbf{P$\checkmark$ (head-to-head)}\\
transient magnitude & P$\checkmark$ (incr.) & tie/neg & P$\checkmark$ (incr.)\\
interior amplification & neg (see text) & P$\checkmark$ (incr.) & neg\\
prev-head identity & P$\checkmark$ (incr.) & neg & P$\checkmark$ (incr.)\\
induction-head identity & neg & neg & neg\\
per-head causal criticality & neg & neg & neg\\
\bottomrule
\end{tabular}
\caption{Verdicts under the pre-registered rule. Persistence of the deviation transient at half depth is the
most robust resolvent result (three of three models; on Llama, family P beats family E outright under both
preprocessings, with $R^2$ of 0.45/0.46 against 0.42/0.31). Representative margins for previous-token
identity on GPT-2: $\Delta R^2=+0.128$ with 95\% confidence interval $[+0.095,+0.155]$ under raw features
and $+0.566$ $[+0.394,+0.734]$ under rank-normal features, and the increment does not depend on \Kdev
(removing it leaves P essentially unchanged). The negatives are uniform for induction identity, where
structural features suffice ($R^2\approx0.6$ for family E), and for causal criticality, where every family
fails ($R^2\approx0$ and top-5 retrieval AUROC below 0.5).}
\label{tab:decisive}
\end{table}

Three annotations are owed to the reader. First, the one tautology. For interior amplification the
raw-feature signal is large ($\Delta R^2=+0.67$) but carried entirely by \Kdev itself; with \Kdev excluded,
$R^2$ drops to $0.005$. Since $\supdev\approx K$ is the content of the Kreiss theorem rather than a fact
about transformers, we count this observable as formally negative, while noting that only the resolvent
family addresses it at all (family E reaches $R^2=0.11$). Second, the criticality wall. No summary from any
family, eigenvalue or resolvent, weight-side or activation-side, predicts which heads are causally critical
under ablation, in any model. The companion paper reports the same wall for weight-space summaries. Together
these results support a boundary statement: single-operator summaries identify head class, not head
criticality, and criticality appears to be a property of wiring rather than of any single operator. Third,
preprocessing sensitivity is reported rather than hidden. The dual-preprocessing rule was adopted after
observing instability, it is the strictest available reading, and both variants are always shown; Pythia's
previous-token identity, for instance, passes under rank-normal features but fails under raw features and is
therefore recorded as negative.

% ================================================================ 9
\section{Related work}
\label{sec:related}

\paragraph{Pseudospectra and non-normality.} The mathematical toolkit is classical
\citep{trefethen2005spectra,trefethen1999computation}, and Kreiss-type constants can be computed exactly
\citep{mitchell2020kreiss}. Non-normal transients serve as a computational resource in recurrent network
design \citep{kerg2019nnrnn} and in neuroscience \citep{murphy2009balanced,goldman2009memory}. Recent
deep-learning applications target training dynamics, namely optimizer update operators
\citep{ghosh2026nonnormal} and optimization Jacobians \citep{ariq2026pseudospectral}, and
\citet{fernando2026dynamics} study non-normality of residual-stream Jacobians with Schur and eigenvalue
tools. To our knowledge, no prior work computes pseudospectra or Kreiss constants of the attention operator
itself, and none compares resolvent against eigenvalue features on behavioral observables.

\paragraph{Attention over depth.} \citet{dong2021attention} prove asymptotic rank collapse of pure attention
stacks. \citet{geshkovski2023clusters,geshkovski2023perspective} model attention as interacting-particle
dynamics with metastable clustering. \citet{naderi2024gap} analyze the spectral gap of the attention matrix
under random-matrix assumptions, and attention rollout \citep{abnar2020rollout} popularized the
depth-product object for attribution. These treatments operate at the level of eigenvalues or asymptotics;
the present contribution concerns the transient regime, with matched nulls, an exact deflation, and the
finding that the eigenvalue heuristic fails by many orders of magnitude specifically on trained stacks.

\paragraph{Sinks and massive activations.} Attention sinks were operationalized by
\citet{xiao2024streamingllm}; \citet{gu2024sinkformation} characterize when they emerge during training;
\citet{guo2024activedormant} describe active and dormant head states; \citet{sun2024massive} identify
massive activations and argue that they act as fixed attention biases; and \citet{barbero2025sinks} link
sinks to compression valleys. We add the operator-level function of sinks as transient dampers, the
training-time ordering in which the pattern never lags the function, the rejection of the overshoot
alternative, and a causal, architecture-conditional localization of the mechanism.

% ================================================================ 10
\section{Discussion}
\label{sec:discussion}

\paragraph{Verdict.} Under the pre-registered rule, the answer to the central question is affirmative only
on a characterized subdomain. The resolvent view is required for identifying learned routing structure in
\A, where the eigenvalues of a triangular matrix are degenerate and classical mixing coefficients can be
vacuous, and for predicting depth-transient survival and persistence, where eigenvalue reasoning errs by
seven to eleven orders of magnitude on trained networks. It is redundant for the static scoring operator
(as the companion paper's tools suffice there), it is unnecessary for induction-head identity (structural
statistics suffice), and it is blind, along with every other single-operator summary we tested, to per-head
causal criticality.

\paragraph{Defaults, reserves, and dampers.} Together with the companion study of the static operator \M, a
consistent organization emerges. Training suppresses the generic spectral structure that masks and random
initialization supply for free, rotational structure in weight space and transient reserve in activation
space, and retains or amplifies that structure only within a small routing subsystem; both suppressions
consolidate after circuit formation on a shared schedule (\S\ref{sec:training}). The two operators express
this organization in complementary algebras. The companion paper shows that position-as-phase makes the
complex eigenvalues of \M informative under \Rope; here the same routing computation appears in \A as
shift-like transport. Rotation in the scoring form and translation in the propagator are the two faces of
one Fourier pair, and the causal mask converts the latter into the textbook non-normal regime of truncated
Toeplitz operators. The two views also differ in what they are invariant to. In the companion's
constrained-training testbed, previous-token function forced through two extreme weight-space
implementations, the free phase-carrying default and a ban-rerouted phase-free alternative whose rotary
phase content differs by two orders of magnitude (rotary fraction $0.003$ to $0.52$), carries positive
\Kdev excess in 16 of 16 heads (minimum $+0.21$, class medians within $0.24$ bits, inside a pre-registered
$\pm0.5$ band). The same contrast replicates in a 160M-parameter, 4B-token real-corpus pilot: all four
previous-token heads are positive (free $+0.36/+0.41$ at the default head positions, rotary fraction
$0.52$--$0.54$; ban-rerouted cosine-only $+0.59/+0.55$ at relocated head positions, rotary fraction near
$0.005$; class medians within $0.19$ bits), against a suppressed-bulk population backdrop in all six runs
(Appendix~\ref{app:pilot}). The weight-space battery reads implementation; the propagator battery reads
function. Rerouting that evades a weight fingerprint, even to the point of relocating the circuit to
different heads, remains visible in the operator dynamics (persistence-type reserve with $\ksup=1$;
two-scale evidence within one training family, with no claim beyond it). Where the sink majority is
concerned, the damper has a physical implementation that the architecture chooses: a narrow
massive-dimension channel on RMSNorm models with full \Rope, causally interceptable as shown in
\S\ref{sec:massive}, and a distributed implementation on LayerNorm models.

\paragraph{Practical readings.} Depth-wise analyses of transformers that rest on spectral gaps or eigenvalue
decay, including signal-propagation and rank-collapse arguments, should be checked against resolvent
quantities, since on trained networks the eigenvalue picture is not a loose bound but the wrong account.
The massive-dimension channel is a single point of failure on Llama-class models: quantization, pruning, or
sparsification that touches those coordinates will damage both function and the model's transient
stabilizer, whereas LayerNorm-class models have no comparably fragile channel, so outlier-aware compression
should be architecture-conditional. The engagement switch provides an operator-level signature of
in-context-learning machinery that is cheap to measure (\Kdev on probe inputs) and is a natural candidate
predictor of in-context ability across checkpoints. For long-context inference, the damping function of
sinks explains why retaining sink tokens stabilizes streaming attention \citep{xiao2024streamingllm}:
evicting them removes the stack's transient governor rather than an attention convention.

\paragraph{Limitations.} The depth-product object is the attention-only rollout abstraction, with head
means and without OV or MLP transformations. The per-head census treats heads as units although heads share
layers and training. Corpus samples are modest (8 to 32 sequences per state), although the effects are large
and replicate. The interior-amplification observable is nearly tautological given Kreiss theory and is
scored as negative. The intervention clamps whole channels, which is stronger than pointwise edits. The
input- and adversarial-sensitivity anchor contemplated in our pre-registration was not executed, so the
connection between transient reserves and adversarial robustness remains open. Finally, both this paper and
its companion find their headline mechanisms to be architecture-conditional; any reading that drops the
conditionality would be an overclaim.

\section*{Reproducibility}
All analyses are in float64 with fixed seeds; censuses, nulls, and contest protocols are deterministic given
the released code, corpus indices, and model revisions. Code, per-head tables, and the full running log with
all pre-registrations and their outcomes, including every negative result, will be released together with
the companion paper.

\bibliographystyle{plainnat}
\bibliography{references}

@misc{fernando2026dynamics,
  title  = {Dynamics of the Transformer Residual Stream: Coupling Spectral Geometry to Network Topology},
  author = {Fernando, Jesseba and Guitchounts, Grigori},
  year   = {2026},
  note   = {arXiv:2605.14258 --- PREPRINT, VERIFY}
}

@book{trefethen2005spectra,
  author={Trefethen, Lloyd N. and Embree, Mark},
  title={Spectra and Pseudospectra: The Behavior of Nonnormal Matrices and Operators},
  publisher={Princeton University Press}, year={2005}
}

@article{trefethen1999computation,
  author={Trefethen, Lloyd N.},
  title={Computation of pseudospectra},
  journal={Acta Numerica}, volume={8}, pages={247--295}, year={1999}
}

@inproceedings{dong2021attention,
  author={Dong, Yihe and Cordonnier, Jean-Baptiste and Loukas, Andreas},
  title={Attention is not all you need: pure attention loses rank doubly exponentially with depth},
  booktitle={ICML}, year={2021}, note={arXiv:2103.03404}
}

@inproceedings{geshkovski2023clusters,
  author={Geshkovski, Borjan and Letrouit, Cyril and Polyanskiy, Yury and Rigollet, Philippe},
  title={The emergence of clusters in self-attention dynamics},
  booktitle={NeurIPS}, year={2023}, note={arXiv:2305.05465}
}

@article{geshkovski2023perspective,
  author={Geshkovski, Borjan and Letrouit, Cyril and Polyanskiy, Yury and Rigollet, Philippe},
  title={A mathematical perspective on transformers},
  journal={Bulletin of the American Mathematical Society}, volume={62}, pages={427--479}, year={2025}, note={arXiv:2312.10794}
}

@inproceedings{naderi2024gap,
  author={Nait Saada, Thiziri and Naderi, Alireza and Tanner, Jared},
  title={Mind the gap: a spectral analysis of rank collapse and signal propagation in attention layers},
  booktitle={ICML}, year={2025}, note={arXiv:2410.07799}
}

@inproceedings{abnar2020rollout,
  author={Abnar, Samira and Zuidema, Willem},
  title={Quantifying attention flow in transformers},
  booktitle={ACL}, year={2020}, note={arXiv:2005.00928}
}

@inproceedings{xiao2024streamingllm,
  author={Xiao, Guangxuan and Tian, Yuandong and Chen, Beidi and Han, Song and Lewis, Mike},
  title={Efficient streaming language models with attention sinks},
  booktitle={ICLR}, year={2024}, note={arXiv:2309.17453}
}

@inproceedings{gu2024sinkformation,
  author={Gu, Xiangming and Pang, Tianyu and Du, Chao and Liu, Qian and Zhang, Fengzhuo and Du, Cunxiao and Wang, Ye and Lin, Min},
  title={When attention sink emerges in language models: an empirical view},
  booktitle={ICLR}, year={2025}, note={arXiv:2410.10781}
}

@misc{guo2024activedormant,
  author={Guo, Tianyu and Pai, Druv and Bai, Yu and Jiao, Jiantao and Jordan, Michael I. and Mei, Song},
  title={Active-dormant attention heads: mechanistically demystifying extreme-token phenomena in {LLMs}},
  year={2024}, note={arXiv:2410.13835}
}

@inproceedings{sun2024massive,
  author={Sun, Mingjie and Chen, Xinlei and Kolter, J. Zico and Liu, Zhuang},
  title={Massive activations in large language models},
  booktitle={COLM}, year={2024}, note={arXiv:2402.17762}
}

@misc{barbero2025sinks,
  author={Queipo-de-Llano, Enrique and Arroyo, {\'A}lvaro and Barbero, Federico and Dong, Xiaowen and Bronstein, Michael and LeCun, Yann and Shwartz-Ziv, Ravid},
  title={Attention sinks and compression valleys in {LLMs} are two sides of the same coin},
  year={2025}, note={arXiv:2510.06477}
}

@inproceedings{kerg2019nnrnn,
  author={Kerg, Giancarlo and Goyette, Kyle and Puelma Touzel, Maximilian and Gidel, Gauthier and Vorontsov, Eugene and Bengio, Yoshua and Lajoie, Guillaume},
  title={Non-normal recurrent neural network ({nnRNN}): learning long time dependencies while improving expressivity with transient dynamics},
  booktitle={NeurIPS}, year={2019}, note={arXiv:1905.12080}
}

@article{murphy2009balanced,
  author={Murphy, Brendan K. and Miller, Kenneth D.},
  title={Balanced amplification: a new mechanism of selective amplification of neural activity patterns},
  journal={Neuron}, volume={61}, number={4}, pages={635--648}, year={2009}
}

@article{goldman2009memory,
  author={Goldman, Mark S.},
  title={Memory without feedback in a neural network},
  journal={Neuron}, volume={61}, number={4}, pages={621--634}, year={2009}
}

@misc{ghosh2026nonnormal,
  author={Ghosh, Souvik},
  title={Non-normal spectral signatures of instability in neural network training dynamics},
  year={2026}, note={arXiv:2605.23476}
}

@misc{ariq2026pseudospectral,
  author={Ariq, Ahanaf Hasan},
  title={Pseudospectral bounds for transient amplification in coupled gradient descent},
  year={2026}, note={arXiv:2606.04031; HiLD workshop, ICML 2026}
}

@misc{mitchell2020kreiss,
  author={Mitchell, Tim},
  title={Computing the {Kreiss} constant of a matrix},
  year={2020}, note={arXiv:1907.06537}
}

@misc{gao2020pile,
  author={Gao, Leo and others},
  title={The {Pile}: an 800{GB} dataset of diverse text for language modeling},
  year={2020}, note={arXiv:2101.00027}
}

@misc{li2026qk,
  author={Li, Hengyu},
  title={When the complex spectrum of attention does work: architecture-conditional non-{Hermitian} structure in the {QK} operator},
  year={2026}, note={Companion paper, arXiv:2607.06621}
}

\appendix
\section{Validation of the $\sigma_{\min}$ core}
\label{app:validation}
\begin{figure}[h]
\centering
\includegraphics[width=.45\linewidth]{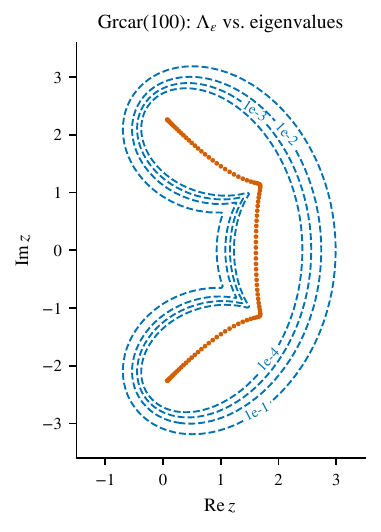}
\caption{Textbook validation: $\Lambda_\eps$ of the Grcar matrix ($n{=}100$) for $\eps$ from $10^{-1}$ to
$10^{-4}$, reproducing the classical contours far beyond the spectrum.}
\label{fig:grcar}
\end{figure}

The grid core reproduces the Grcar pseudospectrum (Fig.~\ref{fig:grcar}) and the Jordan-block $\eps^{1/n}$
law ($\rho_\eps=0.253$ against $\eps^{1/n}=0.251$ at $n{=}10$, $\eps=10^{-6}$). It agrees with dense SVD to
$10^{-6}$ relative error on random, Grcar, Jordan, and stochastic-shift matrices, with comparisons
restricted to the float64-resolvable regime $\sigma_{\min}>10^{-12}\cdot\mathrm{scale}$. It recovers
$K=1.0000$ for normal contractions, verifies the two-sided Kreiss inequality $K\le\sup_k\|A^k\|\le enK$ on
transient test matrices, and verifies Propositions~\ref{prop:sqrtn} and~\ref{prop:deflate} numerically.
Low-rank weight-space operators require an exact $2k$-dimensional orthogonal compression; the frequently
used $k\times k$ core preserves nonzero eigenvalues but not pseudospectra, as the counterexample
$M=e_1e_2^\top$ shows.

\section{Amplifier-head transient modes}
\label{app:amplifier}
Two GPT-2 heads, L1H10 and L11H8, are the only ones in the census whose deflated deviation dynamics
\emph{grows} before it decays: $\|\Adev^k\|_2$ rises with $k$ to an interior maximum at $\ksup=19$ and
$64$ respectively, reaching $15.6$ and $15.9$, so that the Kreiss lower bound (Section~\ref{sec:setup}) is
nearly attained. No matrix from either null family reproduces this behavior; all nulls have
$\ksup\equiv1$, meaning that their deviation dynamics contracts from the first step.

\begin{figure}[h]
\centering
\includegraphics[width=.95\linewidth]{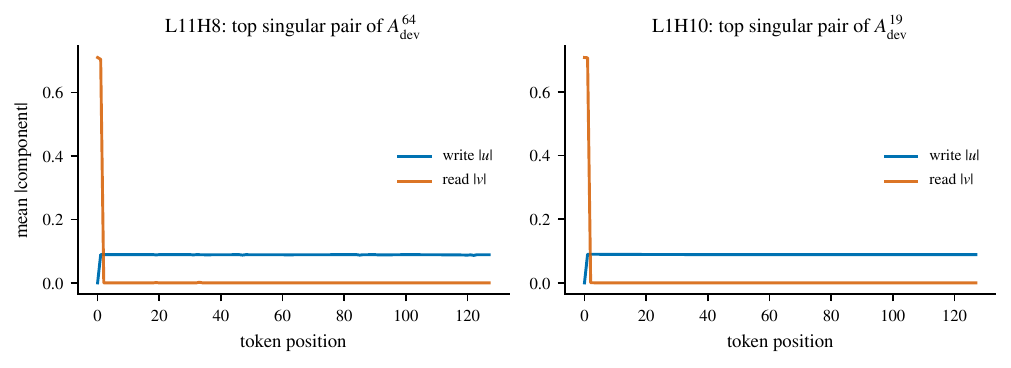}
\caption{Top singular pair of $\Adev^{\ksup}$ for the two interior amplifiers (GPT-2 L11H8 and L1H10),
averaged over sequences. The transient reads the contrast between BOS and the first content token
($v\approx(e_0\pm e_1)/\sqrt2$) and writes it nearly uniformly across downstream positions
($|u|\approx n^{-1/2}$).}
\label{fig:amplifier}
\end{figure}

To characterize what is amplified, we take, for each corpus sequence, the singular value decomposition of
$\Adev^{\ksup}$ and retain the leading singular pair $(u,v)$. The right vector $v$ is the input direction
that the transient amplifies most, the left vector $u$ is where that amplification is written, and the
leading singular value is $\|\Adev^{\ksup}\|_2$. Because singular vectors are defined up to sign, we
average $|u|$ and $|v|$ componentwise across sequences; the profiles are stable, with the spread across
sequences smaller than the line width in Fig.~\ref{fig:amplifier}.

The resulting mode is simple and identical in both heads (Fig.~\ref{fig:amplifier}). The read direction is
concentrated on the first two positions, $v\approx(e_0\pm e_1)/\sqrt2$, which is a contrast between the
BOS token and the first content token; every other coordinate is below $0.01$. The write profile is nearly
flat, $|u|\approx n^{-1/2}$ across all downstream positions. The two heads therefore act as broadcast
amplifiers of an early-context contrast: they extract one scalar from the beginning of the sequence and
distribute it, amplified by a factor of order $\ksup$ applications of the deflated operator, over the whole
context. This is the only structure in our census for which the interior-growth regime of Kreiss theory is
realized by a trained attention head, and it is invisible to the eigenvalues of the head, which are its
diagonal and lie near zero.

\section{The 160M real-corpus implementation-invariance pilot}
\label{app:pilot}
Six Pythia-160M-architecture models (12 layers $\times$ 12 heads, $d{=}768$, rotary fraction $0.25$) were
trained from scratch on 4B tokens of real corpus under the companion paper's constrained-training protocol
\citep{li2026qk}: two free seeds; two seeds with the companion's rotary-phase ban regularizer
($\lambda{=}10$), under which the previous-token circuit re-forms as a cosine-only implementation at
different head positions; and two scaffolded seeds, which are descriptive here. Implementation was read from
the weights with the companion's rotary-fraction metric, and transient reserve with this paper's battery,
under a protocol identical to the main censuses (Perron-deflated \Kdev against per-matrix row-permutation
nulls, the same fixed 16-sequence $n{=}128$ corpus batch and seed formulas, and bfloat16 capture upcast and
gate-checked to $|\Delta|\le0.002$ against the training-side evaluation). All four previous-token heads
carry positive paired excess: free $+0.361$ and $+0.413$ (heads L2H0 and L2H5, rotary fraction
$0.52$--$0.54$) against ban-rerouted $+0.594$ and $+0.552$ (relocated heads L5H8 and L3H0, rotary fraction
$0.0046$ and $0.0059$), with class medians $+0.387$ and $+0.573$ ($|\Delta|=0.186$, inside the
pre-registered $\pm0.5$ band), and with the suppressed-bulk backdrop present in every run (population median
excess of the 143 non-prev heads between $-0.15$ and $-0.73$; 58\% to 76\% of heads below zero). Scope:
two seeds per class from a pilot inventory; 4B-token models, which are formation-complete but far from
convergence; persistence-type reserve ($\ksup\equiv1$) rather than the interior-amplification type of
\S\ref{sec:census}; one fixed evaluation batch. No generalization beyond this training family is claimed.
The formation-time dynamics of the same runs are analyzed in a separate note in preparation; no result from
that note is relied upon here.

\section{Pseudospectra of real heads against their nulls}
\label{app:demo}

\begin{figure}[!ht]
\centering
\includegraphics[width=.80\linewidth]{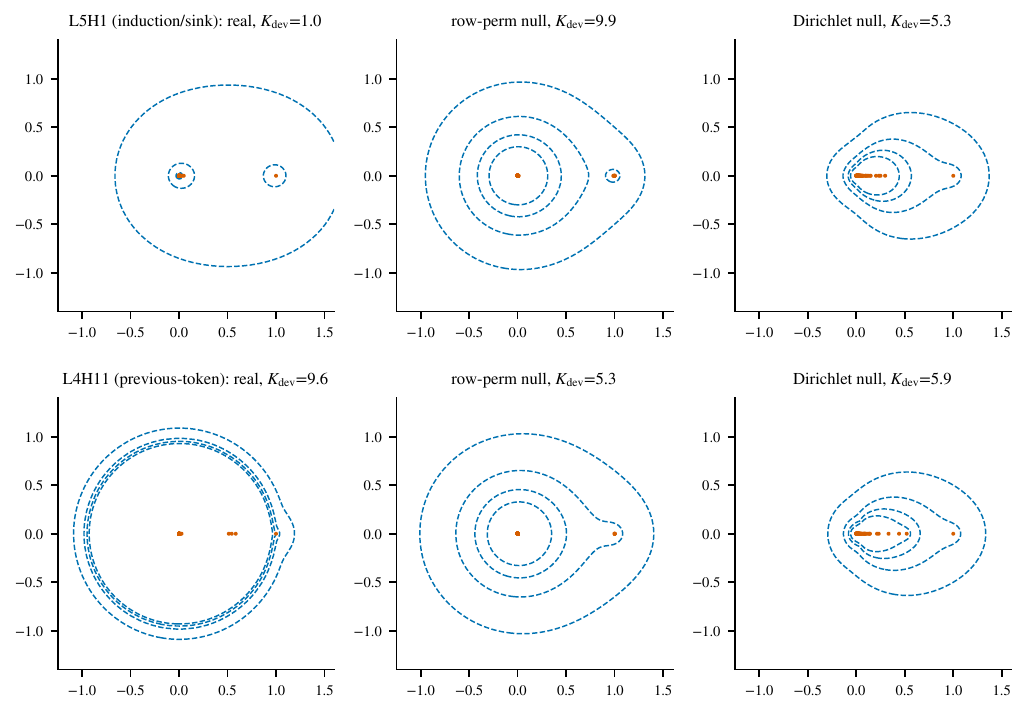}
\caption{$\Lambda_\eps$ contours ($\eps$ from $10^{-1}$ to $10^{-4}$) for one input sequence: an
induction/sink head (top row) and the previous-token head 4.11 (bottom row) against their row-permutation
and Dirichlet nulls. The previous-token head shows the truncated-Toeplitz ring near $|z|=1$; the sink head
collapses to near-normal structure, with $\Kdev=1.0$, below its null.}
\label{fig:demo}
\end{figure}

The scalar summaries used throughout the paper are level-set statistics of the resolvent norm. It is
instructive to see the underlying sets. Figure~\ref{fig:demo} shows $\Lambda_\eps$ for two GPT-2 heads on a
single input sequence, each against its row-permutation and Dirichlet nulls, computed on the undeflated
matrices so that the mask-forced Perron eigenvalue at $z=1$ is visible.

The two heads illustrate the two ends of the signed census. The previous-token head 4.11 has a
pseudospectrum that forms a ring near $|z|=1$: its $\eps$-level sets enclose the unit circle even though its
eigenvalues, which for a triangular matrix are its diagonal, consist of the mask-forced value $1$ together
with a cluster near the origin. This is the classical
picture of a truncated Toeplitz shift, whose spectrum collapses under truncation while its pseudospectrum
retains the symbol curve \citep{trefethen2005spectra}, and it is the operator-level expression of what the
head computes, namely near-deterministic transport by one position. The induction head 5.1, dormant on this
generic-text input, is at the opposite end: its $\eps$-level sets are close to the $\eps$-balls of a normal
operator, its deflated Kreiss constant is $1.0$, and it therefore sits below its own mass-matched null.

Two cautions attach to this figure. The annotated constants are those of the single matrices shown, whereas
the census statistics of Section~\ref{sec:census} are per-head medians over sequences, each paired with its
own null draw; for these two heads the medians are $9.63$ against $9.12$ (head 4.11) and $1.00$ against
$8.64$ (head 5.1). The panels therefore illustrate geometry rather than effect size. The row-permutation
null can reach comparable Kreiss values, as the head-4.11 comparison in Section~\ref{sec:census} makes
explicit, yet its level sets remain simply connected and centered on the origin. The ring, and the
near-normal collapse of the sink head, are learned structure.

\end{document}